\documentstyle [12pt]{article}
\def\be{\begin {equation}}
\def\ee{\end {equation}}
\def\bea{\begin{eqnarray}}
\def\eea{\end{eqnarray}}
\def\frac#1#2{{\displaystyle#1\over\displaystyle#2}}
\def\d{\partial }
\def\o{\omega }
\def\kappa{\ae}
\def\br{({\bf r})}
\def\ot{\tilde{\omega}}
\def\rl{({\bf r}, \lambda)}
\def\m{^{(m)}}
\def\sqr{\sqrt}
\def\intl{\int \limits}

\def\l{\lambda}

\begin {document}

\author{ G.A. El \footnotemark[1]}

\date{Max-Planck-Institute for Aeronomy, D - 37191 Katlenburg-Lindau,
Germany}

\title{ On evolution of multiphase nonlinear modulated waves }

\maketitle

\begin{abstract}
We present a fundamental solution to an initial value problem for the
KdV-Whitham system in an explicit integral form. Monotonically decreasing
initial data with finite number of breaking points are considered. Generating
function for the commuting flows of the averaged KdV hieararchy producing the
analytical solutions to the KdV-Whitham system is constructed.
\end{abstract}

\footnotetext[1]{On leave of absence
from Institute of Terrestrial Magnetism, Ionosphere
and Radio Wave Propagation, 142092, Troitsk, Moscow Reg., Russia}

\section {Introduction}

It is known [1,2] that  evolution of smooth initial distribution
\be
u(x,0) = u_0 (x)    \label{a1}
\ee
according to the Korteweg-de Vries (KdV) equation
\be
\partial _t u + 6u \d _x u + \varepsilon ^2 \d ^3_{xxx} u = 0  \label{a2}
\ee
is described locally as $\varepsilon \rightarrow 0 $ by
$g$ - phase ($g$ - gap)
solution which is expressed in terms of theta-function of the
hyperelliptic Riemann surface of genus $g$ (see Refs. [3,4])
\bea
\Gamma \; : \ \  && y^2 = \prod \limits _{j=1} ^{2g+1} (\lambda - r_j)
\equiv R_{2g+1} ({\bf r}, \lambda) ,       \label{a3} \\
&& r_1 \le r_2 \le ... \le r_{2g+1} .  \nonumber
\eea
This solution has the form
\be
u(x,t) = -2 \frac{d^2}{dx^2} \ln \theta
\left( \left. \frac{{\bf k}x-{\bf \omega}t}{\varepsilon} \right|
r_1, ... , r_{2g+1} \right) + c({\bf r}) \ .    \label{a4}
\ee
Here the components of the wave number and the frequency vectors
$ {\bf k} \br = \left( k^{(1)}, ... , k^{(g)} \right) \;$,
$ {\bf \omega} \br = \left( \o^{(1)}, ... , \o^{(g)} \right) $
are  expressed in terms of the branch points $r_j$ by well-known
"finite-gap" formulas
\be
k^{(m)} = \kappa _1^{(m)} \; , \ \
\o^{(m)} = \kappa _1^{(m)} 2 \sum _{j=1}^{2g+1} r_j + 4 \kappa _2^{(m)} \ ,
\ \ m=1,...,g \ .
\label{a5}
\ee
where the functions $\kappa_i^{(m)} ({\bf r})$ are defined uniquely by the
normalization conditions
\be
\oint \limits _{\alpha_k} d\Omega ^{(m)} = \delta _{km} \ , \  \ \
k, m = 1,...,g       \label{a6}
\ee
for the basis    holomorphic differentials
\be
d \Omega ^{(m)} = \sum _{k=1}^g \kappa _k^{(m)} ({\bf r})
\frac{\lambda ^{g-k}}{\sqrt{R_{2g+1}({\bf r},\lambda)}} d\lambda \ , \label{a7}
\ee
where the sign is given by $\sqr{1}=1$.

We consider the Riemann surface (3) with cuts along the intervals
$(-\infty,r_1], [r_2,r_3]$, ...,$ [r_{2g}, r_{2g+1}]$.
The canonical basis of cycles $\alpha_1,...,\alpha_g ; \ \;
\beta_1,...,\beta_g$ is chosen so that the cycles
$\alpha_1,...,\alpha_g $
lie on $\Gamma $ over the permitted zones $[r_1,r_2],...,[r_{2g-1},r_{2g}] $.
The genus $g$ of the Riemann surface depends on the initial
data $u_0(x)$ (globally) and on the position on
the $(x,t)$ -- plane (locally).
In the general case the  $(x,t)$ -- plane is splitted into a finite
set of regions in every of which the solution is descibed by the
formula ~(\ref{a4}) with certain $g$ [5,6]  (Fig. 1). The slow
$x,t$ -- evolution of the branch points $r_j$ parametrizing the
"uniform" solution ~(\ref{a4}) is governed by the Whitham modulation
equations [7,8,1] and provides the required transition. These
equations have a diagonal (Riemann) form
\be
\d _t r_i + V_i({\bf r}) \d _x r_i =0 \ , \ \ i=1,2,...,2g+1  \label{a8}
\ee
with real and different characteristic speeds (pure hyperbolicity
[9])  which will be specified below.
The existence of global solutions to ~(\ref{a8}) corresponding to the
monotonically decreasing analytic initial data ~(\ref{a1}) for the
KdV equation was proven recently by Tian in Ref. [10] where a set of
analytic solutions in the form of infinite series were constructed with
the aid of the generalized hodograph transform [11] and Krichever
algebraic -- geometrical procedure [12].

Unlike that of Ref. [10] we propose the direct way to construct the general
(nonanlytic) global solutions of ~(\ref{a8}) in an explicit integral form. This
 problem
has been solved to date only in the simplest single--phase case [13-16].
Construction of the global solutions for the multiphase case
is vital both from purely
theoretical point of view to complete the theory of the multi-gap averaged
integrable systems and for the physical applications where the complex
multiphase
nonlinear dissipationless wave behaviour is not uncommon
(Alfvenic turbulence in the solar wind [17], the Earth's
bowshock [18] and oth.)

\section{ Generating function for the averaged  KdV    hierarchy }

We start from the potential representation for the characteristic
speeds $V_j$ :
\be
V_j ({\bf r}) = \frac{\d _j \o ^{(m)}}{\d _j k^{(m)}} \ ; \ \ \ \
\d _j \equiv  \d / \d r_j     \label{a9}
\ee
which immediately follows from the "conservation of waves"
\be
\d _t {\bf k} ({\bf r}) + \d _x {\bf \o } ({\bf r}) = 0 \ ,  \label{a10}
\ee
considered as a consequence of ~(\ref{a8}) (for the single--phase case see
Refs.[14,15,19]).
We notice that the representation (\ref{a9}) is valid for any chosen phase
number since the characteristic speeds $V_j$ do not depend on $m$.

Looking for the solution of ~(\ref{a8}) in the characteristic form
\be
x - V_i ({\bf r}) t = W_i ({\bf r})  \label{a11}
\ee
one arrives at the linear overdetermined compatible equation system
for $W_i$'s  [11,20].
\be
\frac{\d _i W _j}{W_i - W_j} = \frac{\d _i V _j}{V_i - V_j} \ ;\ \ \ \
i \ne j \ .
\label{a12}
\ee
Construction ~(\ref{a11}), ~(\ref{a12}) is known as the generalized
hodograph transformation and provides all the hydrodynamic symmetries of
{}~(\ref{a8}), i.e. equations of the form
\be
\d _{\tau} r_i + W_i \br \d _x r_i = 0 \ ,    \label{a13}
\ee
which commute with ~(\ref{a8}), i.e.
$ \d _{t \tau} r_i = \d _{\tau t} r_i $.
Here $\tau $ is "new" independent time. Some of the solutions of ~(\ref{a12})
(the uniform ones) correspond to the averaged hierarchy of the KdV equation
[8,12,20,21].
The system ~(\ref{a12}) can be locally parametrized by $2g+1$ functions of one
variable [20]. We remark that this corresponds to the number of arbitrary
functions $r_j (x,0)$ in the initial value problem for the Whitham system
{}~(\ref{a8}).

Now we shall construct a generating function $W_j({\bf r}; \lambda)$ for the
uniform commuting flows to ~(\ref{a8}) which will give the characteristic
speeds of the averaged KdV hierarchy as the coefficients of the expansion
of $W_j({\bf r}; \lambda)$ in powers of $\lambda ^{-1}$ as
$\lambda \rightarrow \infty $.
The two first terms of this series are obvious and correspond to the
zeroth (linear equation of translation) and the first (system ~(\ref{a8}))
terms of the averaged KdV hierarchy:
\be
W_j^0 = 1 \ , \ \ \ W_j^1 = V_j \ .   \label{a14}
\ee
So we shall seek $W_j({\bf r}; \lambda)$ in the form predicted by ~(\ref{a9}),
{}~(\ref{a12}):
\be
W _j ({\bf r};\lambda) = \frac{\d _j \ot ^{(m)} ({\bf r}; \lambda)}
{\d _j k^{(m)}\br} \ ,  \ \  m=1,...,g,    \label{a15}
\ee
where
\be
\ot ^{(m)} ({\bf r}; \lambda) = k^{(m)} + \frac{const}{\lambda}
\o ^{(m)} + ...   \label{a16}
\ee
Such a generating function for the single--phase case was constructed by Pavlov
in [22] and has the form
\be
\ot _{s.p.} ({\bf r};\lambda) = \frac{k \lambda ^{3/2}}{\sqr{R_3 \rl}}\ .
\label{a17}
\ee
With ~(\ref{a15}), ~(\ref{a16}), ~(\ref{a17}) in mind we propose the following
project:
\be
\ot ^{(m)} ({\bf r}; \lambda) = \lambda ^{3/2} \sum _{k=1} ^g \kappa _k \m
\frac{\lambda ^{g-k}}{\sqr{R_{2g+1}\rl}} =
\lambda ^{3/2} \frac{d \Omega \m}{d\lambda} \ ,    \label{a18}
\ee
which has two first terms of expansion as $\lambda \rightarrow \infty $
coinciding with ~(\ref{a16}) and turns into ~(\ref{a17}) in the case
$g=1$.

Let us check that commuting flows ~(\ref{a15}) with the frequences
{}~(\ref{a18}) identically satisfy the linear equation system ~(\ref{a12}).

With this aim in view we consider the following combinations occuring in the
left part of ~(\ref{a12}):
\be
\lambda ^{-3/2} \d _i
\left( \frac{\d _j \ot \m}{\d _j k \m} \right) =
p_{ij}\m \br \frac
{\lambda^g + a_{g-1}\m \lambda^{g-1}+...+a_0\m}
{(\lambda -r_j)(\lambda-r_i) \sqr{R_{2g+1}\rl}} \ , \label{a19}
\ee
and
\be
\lambda ^{-3/2}
\left( \frac{\d _i \ot \m}{\d _i k \m} -
 \frac{\d _j \ot \m}{\d _j k \m} \right) =
q_{ij}\m \br \frac
{\lambda^g + b_{g-1}\m \lambda^{g-1}+...+b_0\m}
{(\lambda -r_j)(\lambda-r_i) \sqr{R_{2g+1}\rl}} \ . \label{a20}
\ee
Here $p_{ij}\m \br$, $q_{ij} \m \br$ are some certain (analytic) functions.
The coefficients $a_k \m \br$, $b_k \m \br$ can be uniquely defined by the
normalization conditions, namely, integrating ~(\ref{a19}), ~(\ref{a20})
over the $\alpha$--cycles we arrive taking into account ~(\ref{a6}), (18) at
two
sets of algebraic equations
\bea
&& \oint \limits _{\alpha _k} \frac
{\lambda^g + a_{g-1}\m \lambda^{g-1}+...+a_0\m}
{(\lambda -r_j)(\lambda-r_i) \sqr{R_{2g+1}\rl}} d\lambda = 0 \ , \nonumber \\
&& \hbox{and}   \label{a21}  \\
&& \oint \limits _{\alpha _k} \frac
{\lambda^g + b_{g-1}\m \lambda^{g-1}+...+b_0\m}
{(\lambda -r_j)(\lambda-r_i) \sqr{R_{2g+1}\rl}} d \lambda = 0 \ ,
\ \ k=1,...,g \ ,   \nonumber
\eea
which are identical and imply $a_j \m \br = b_j \m \br ,\ j=0,1...,g-1 $.
Therefore the ratio
\be
S_{ij} \rl = \frac
{\d _i \left( \frac{\d _j \ot \m}{\d _j k \m} \right)}
{\frac{\d _i \ot \m}{\d _i k \m} - \frac{\d _j \ot \m}{\d _j k \m} } =
\frac{p_{ij} \m \br}{q_{ij} \m \br} \ ;    \label{a22}
\ee
does not depend on $\lambda$ and to check the fulfilment of the equality
{}~(\ref{a12}) for the functions $W_j (\lambda; {\bf r})$ defined by
{}~(\ref{a15}), ~(\ref{a18}) it is suffice to check it for any of terms of the
expansion ~(\ref{a16}). The equation is obviously satisfied by the second
term which gives the simplest nontrivial solution $W_j = V_j$.
Hence the function ~(\ref{a18}) really is the generating function for
the frequences of the KdV hierarchy. As a consequence, the generating equation
for  the phase wave number conservation laws of the KdV hierarchy has the form
\be
\d _{\tau} k \m + \d _x \left( \frac{d \Omega \m}{d \lambda}
\lambda^{3/2} \right) = 0 \ .   \label{a23}
\ee

\section{ General solution of the hodograph equations}

Now we are ready to construct the general solution to the hodograph equations
{}~(\ref{a12}). Remind that it has to contain $2g+1$ arbitrary functions of one
variable. So using the generating function $W_j \rl $ as a kernel of
convolution we arrive at the general solution in the
form
\be
W_j \br = \sum _{k=1}^{2g+1} \oint \limits _{A_k} W_j ({\bf r; \lambda})
\phi _k (\lambda) d
\lambda \ , \label{a24}
\ee
where $\phi _k (\lambda)$ and $A_k$ are arbitrary
functions and contours respectively on the Riemann surface $\Gamma$. The
important restriction that has to be imposed upon $\phi _k$'s and $A_k$'s
follows from the pure hyperbolicity of the modulational system ~(\ref{a8}) [9]
and the form of the hodograph solution ~(\ref{a11}), namely, all $W_j$'s must
be real. We remark that this requirement is not contained in the system
{}~(\ref{a12}) itself. Moreover, the contours $A_k$ have to lie off the
branch points. Finally, using the "potential" representation
{}~(\ref{a15}), ~(\ref{a19}) we arrive at the solution
\be
W_j \br = \frac {\d _j
\ot \m \br}{\d _j k \m \br} \ , \ \ j=1,2,...,2g+1, \ \ m=1,...,g,
\label{a25a}
\ee
where
\be
\ot \m \br = \sum _{k=1} ^{2g+1} \intl _{a_k} ^{r_k}
\phi _k (\lambda) d \Omega \m \ .  \label{a25b}
\ee
Here "new" arbitrary $\phi_k$'s correspond to $\lambda ^{3/2} \phi _k
(\lambda)$
from ~(\ref{a24}). Arbitrary constants $a_k$ are real therefore the integration
is
accomplished along the axes in  real $r$--space (this corresponds to a special
choice of $A_k$'s which can be tightened to a real axis on the Riemann
surface).
Choosing
$\phi_k(\lambda)\;, a_k$ one can obtain all the commuting flows described in
[21,23].

Now we shall discuss briefly connection between the solution ~(\ref{a25a}) and
the Euler--Darboux--Poisson  type systems which were investigated much
earlier by Eisenhart [24] for the case of three
independent variables ($g = 1$) and  have
arisen recently when studying the finite--gap averaged KdV equation
[25,14,15,16,10].
In fact, the solution ~(\ref{a25b}) can be represented in the form
\be
\ot \m \br = \sum _{i=1} ^g  \kappa _i \m \br f_i \br  \label{a26}
\ee
where $\kappa _i \br$ are the normalization coefficients defined by
{}~(\ref{a6}),
{}~(\ref{a7})  and
\be
f _i \br = \sum _{k=1} ^{2g+1} \intl _{a_k} ^{r_k}
\frac{\lambda ^{g-i} \phi _k (\lambda)}{\sqr{R_{2g+1} \rl}} d \lambda \ ,
\ \ i=1,...,g    \label{a27}
\ee
are the  Eisenhart type solutions of the overdetermined compatible system
of the Euler--Darboux--Poisson equations
\be
2 (r_i - r_j) \d _{ij} ^2 f = \d _i f - \d _j f \ , \ \ i\ne j \ , \ \
i, j = 1,...,2g+1 \ . \label{a28}
\ee
The solution ~(\ref{a25b}) generally is unlimited when any two of neighboring
invariants $r_j$ coalescing. For the further consideration we shall need
the solution which is limited in all the domain of definition
$r_1 \le ... \le r_{2g+1} $.
The requirement of boundedness reduces the number of arbitrary functions in
the general solution. We first demonstrate the way to construct
the limited solution for the simplest single--phase case (see Refs. [14,15]).
We start from the representation ~(\ref{a25b}) which takes in the
single--phase case the form
\bea
&& \ot \br = \sum _{k=1} ^3 \intl _{a_k}^{r_k} \phi _k (\lambda) d \Omega \ ,
\nonumber \\
&& \label{a29} \\
&& d \Omega = \frac {k}{\sqr{R_3 \rl}} d \lambda \ ,  \ \
\intl_{r_1}^{r_2} d \Omega = 1 \ .  \nonumber
\eea
To provide boundedness of ~(\ref{a29}) at the singular surfaces $r_2=r_1$
and $r_2=r_3$  one splits the domain of definition $D$: $r_1 \le r_2 \le r_3$
into two parts $D_1: \ \xi < r_2 \le r_3$ and $D_2: \ r_1 \le r_2 <\xi $
with the deleted point (surface)
$r_2=\xi$, where $r_1 \le \xi \le r_3$ is some real
parameter on the Riemann surface. Each of these subdomains contains only the
one singularity. To
obtain the limited solution one should put $a_1 = a_2 = a_3 = \xi $ and
\bea
&&\phi _2 = - \phi _3 \ \ \ \ \ \hbox{for} \ \ D_1 \ , \nonumber \\
&& \label{a30} \\
&&\phi _2 = - \phi _1 \ \ \ \ \ \hbox{for} \ \ D_2  \ . \nonumber
\eea
By this means the resulting solution is characterized by two (instead of
initial three) arbitrary functions $\phi _1(\lambda)$ and $\phi _3 (\lambda)$.
Introducing new (real) arbitrary functions $\psi _1 (\lambda) = - \phi _1
(\lambda) \; ,
\psi_2 (\l) = \sqr{-1} \phi _3 (\l) $ we present the general bounded continuous
solution in the form
\bea
&& \intl _{r_1} ^{r_2} \psi _1 (\l) d \Omega  + \sqr{-1} \intl _{\xi}^{r_3}
\psi_2(\l) d \Omega \ , \ \ \ \hbox{for} \ r_1 \le r_2 \le \xi  \nonumber \\
\ot \br =   &&  \label{31}  \\
&& \intl _{r_1} ^{\xi} \psi _1 (\l) d \Omega  + \sqr{-1} \intl _{r_2}^{r_3}
\psi_2(\l) d \Omega \ , \ \ \ \hbox{for} \ \xi \le r_2 \le r_3  \nonumber
\eea
or more compactly
\be
\ot \br = \intl _{r_1} ^{min(r_2, \xi)} \psi _1 (\l) d \Omega +
\sqr{-1} \intl _{max(r_2, \xi)} ^{r_3} \psi _2 (\l) d \Omega \label{a32} \ .
\ee
The obtained solution can be singular only at two points which are outside the
domains $D_1$ and $D_2$, namely,  $r_2 = r_1 = \xi $ and $r_2 = r_3 =\xi $.
These singularities are removed by the additional requirement
\be
\psi_1 (\xi) = \psi _2 (\xi) = 0 \ .  \label{a33}
\ee
We remark that the solution ~(\ref{a32}), ~(\ref{a33}) generally is nonanalytic
at the plane $r_2 = \xi $.  This, however, does not contradict to a hyperbolic
nature of the problem under consideration. Such a weak singularity was
first revealed by Gurevich and Pitaevskii [13] in the numerical solution for
the
problem of cubic-like breaking (the exact solution of this problem,
however, is analytic and does not contain any
singularities [26,15,27]).  The solution ~(\ref{a32}), ~(\ref{a33}) can be
easily generalized to the multiphase case. The result is:
\bea
&& \ot \m \br =
\sum _{k=1} ^{g} \ot _k \m \br \ , \ \ m=1,...,g,  \nonumber \\
&& \hbox{where}  \label{a34} \\
&& \ot _k \m \br = \intl _{r_{2k-1}} ^{min(r_{2k}, \xi_k)} \psi
_{2k-1} (\l) d \Omega \m + \sqr{-1} \intl _{max(r_{2k}, \xi_k)} ^{r_{2k+1}}
\psi _{2k} (\l) d \Omega \m \ , \nonumber \\
&& \nonumber \\
&& \nonumber \\
&& \psi _{2k-1}  (\xi _k) = \psi _{2k} (\xi _k) = 0 \ ,   \nonumber  \\
&& r_{2k-1} \le \xi_k \le r_{2k+1} \ .  \label{a35}
\eea
One can see that this
solution is continuous and limited in all the domain of definition and
generally contains $g$ weak singular points
$r_{2k} = \xi _k \ , \ \ k=1,...,g $.
We remark in addition that formula ~(\ref{a34}) represents in some sense a
decomposition of the $g$--phase solution into $g$ single-phase ones in the
hodograph space.

The analytic limited solutions which were obtained in [10] in the form
of infinite series can be picked out from the general representation
{}~(\ref{a34}) by imposing the additional restriction
\be
\psi _{2k-1} (\l) = \psi _{2k} (\l) \equiv \psi (\l) \ , \ \ k=1,...,g \ ,
\label{a36}
\ee
where $\psi (\l)$ is a real analytic function with $g$ real zeros $\xi _k$.

\section{Boundary value problem solution}

Now we formulate the boundary problem to the system ~(\ref{a12})
corresponding to the initial value problem ~(\ref{a1}), ~(\ref{a2}).
To avoid geometrical complifications we shall consider only
monotonically decreasing initial data ~(\ref{a1}). Evolution of such
a curve with the only breaking point ( $ u''(x_{br}) = 0 $ ) is described
completely  (in the weak limit sense [1,2,6]) by the single--phase averaged
system ~(\ref{a8}) with the boundary conditions formulated first by
Gurevich and Pitaevskii [13]. The more recent investigations [1,2,16,28]
have shown that these boundary conditions arise naturally in the zero
dispersion limit of the KdV equation and are in the continuity of the
Riemann invariants $r_j$ on the phase transition boundaries where
$r_{2k} = r_{2k-1}$ ($j \ne 2k,2k-1 $) or $r_{2k}=r_{2k+1}$
($j \ne 2k, 2k+1$). Earlier this problem was investigated by Avilov and
Novikov [29] numerically in the framework of the theory of system of
the hydrodynamic type. The effective analytic solutions of the
Gurevich -- Pitaevskii problem for all single--phase evolving (including
nonanalytical and hump--like ) data were obtained in Refs. [15,30,31].
It is clear that every additional breaking introduces a new phase (gap)
into consideration and therefore the genus of the problem $g$ coincides
with the number of breaking points on the initial curve [6]. By this means
the number of dynamical variables in the problem is $2g+1$. At some
moment $t_{overlap}$ the evolving oscillatory phases will begin to overlap
forming the multiphase solution region.
To be specific we consider for the beginning the case when the initial curve
$u_0(x)$ has two distinct breaking points $x_1$ and $x_2$ (Fig. 2a):

\bea
&& u_0'(x)<0 \ , \ \ u_0''(x_1) = 0 \ , \ \ u_0''(x_2)=0\ , \nonumber \\
&& u_0'''(x_1)>0 \ , \ \ u_0'''(x_2)>0 \ , \label{a37} \\
&& u_0(x_1) = c_1 \ , \ \ u_0 (x_2)=c_2 \ . \nonumber
\eea
We assume in addition that the initial data are such that the breakings
appear distinctly in space (i.e. that there are not points like $A$ in
Fig. 1). This assumption is made only for convenience of consideration
and will be omitted afterwards. To obtain boundary conditions to the system
{}~(\ref{a12}) corresponding to the initial data ~(\ref{a1}), ~(\ref{a37})
we apply to the one--phase stage of the evolution when two single-phase
modes do not overlap yet (Fig. 2b). At this stage the global solution
of ~(\ref{a12}), which is generally defined in  five $(2g+1)$ -- dimensional
$r$-space, lies on two $(g)$ separated three--dimensional sections
$(r_1, r_2, r_3), ...,(r_{2g-1}, r_{2g}, r_{2g+1})$  (Fig. 3) (see Ref.[15])
with common diagonal surfaces $r_2=r_3=r_4,..., r_{2g-2}=r_{2g-1}=r_{2g}$.
These diagonal surfaces are the domains of definition of the zero-phase
solution which separates the one-phase domains at the stage considered. Due to
hyperbolicity of the hodograph equations (which is inherited from the
hyperbolicity of the KdV-Whitham system) these 3-D domains do not
interact and can be considered independently.
We remark that the initial data and their smooth evolution until the first
breaking
are defined, as it will be shown below, on the pure diagonal
$r_1=r_2=...=r_{2g+1}$ in hodograph space [16,10].

Now we recall the important general fact proven in [10],
which provides existence of the global solution to the initial value
problem for the KdV-Whitham system: the uniform commuting flows $W_j \br$
(including $V_j \br$, of course) do not depend on multiple invariants
$r_{2k}=r_{2k+1}$, or $r_{2k}=r_{2k-1}$ if $j \ne 2k, 2k+1$ or $2k-1$,
and, moreover, they turn into their analogs for the $g-1$ genus case:
\bea
&& W_j
\left| \begin{array}{l}
 \\  \ \  (r_1,...,r_{2k-1}, \beta, \beta, r_{2k+2}, ..., r_{2g+1})   \\ g \\
 \end{array}  \right.
=  \nonumber \\
\label{a38}  \\
&& W_j
\left| \begin{array}{l}
 \\  \ \  (r_1,...,r_{2k-1}, r_{2k+2}, ..., r_{2g+1}), \\ g-1 \\
 \end{array}  \right.
 \ \ \ j \ne k, k+1 \ ,           \nonumber
 \eea
and the analogous assertion is true for $r_{2k}=r_{2k-1}, \ \; j\ne 2k, 2k-1$.
As the obvious consequence of the equations ~(\ref{a12}) these relationships
take
place for all (not only uniform) $W_j$'s.
This fact allows us to formulate boundary conditions at the phase transition
 surfaces $r_{2k}=r_{2k-1}$ and $r_{2k}=r_{2k+1} , \ \; k=1,...,g$
 which are single-phase -- zero-phase transitions at the considered stage of
 evolution. Comparison of the hodograph solution ~(\ref{a11}) with the
 solution of the Hopf equation (zero-phase averaged KdV)
 \be
 x-6rt = W(r) \ ,  \label{a39}
 \ee
 where $W(r)$ is the inverse function to the initial curve $u_0 (x)$
 ~(\ref{a1}), ~(\ref{a37}) (see Fig. 2a) shows that odd branches of the
 multivalued curve $r_j(x,r), \ j=1,...,5; \ r_1\le ...\le r_5$, have to
 be matched with the appropriate branches of the inverse function $W(r)$
 (see Fig. 2b). We shall demonstrate in detailes how to accomplish this
 matching at the boundaries where $r_2=r_3$ and $r_2=r_1$ (the right-hand
 single-phase wave in Fig. 2b). The rest matching can be accomplished in
 an analogous way.

 As a first step we make a passage to the single-phase regime in the
 two-phase desired solution ~(\ref{a12}) with certain $W_j(r_1,...,r_5)$.
 This corresponds to  passage to times
 $t_{br}<t<t_{overlap}$.

 Using ~(\ref{a38}) we have the transition
 \be
 W_j
 \left| \begin{array}{l}
 \\  \ \  (r_1,r_2,r_3,\zeta, \zeta)   \\ g=2
 \end{array}  \right.
 =
 W_j
 \left| \begin{array}{l}
 \\  \ \   (r_1,r_2,r_3) \\ g=1
 \end{array}  \right.
 \ ,
 \ \ j =1,2,3 \ ,          \label{a41}
 \ee
 where $\zeta$ is an arbitrary real parameter.

 Now the boundary conditions for the function $W_j(r_1, r_2, r_3)$
 can be obtained by means of the usual Gurevich--Pitaevskii matching
 (see Refs. [13-15]) for the single-phase case, namely, at the edges
 where $r_2=r_3$ and $r_2=r_1$ we have, taking into account ~(\ref{a38}),
 the single-phase -- zero-phase transitions which drive us to the solution of
 the Hopf equation ~(\ref{a39}) with the initial data ~(\ref{a1}),~(\ref{a37})
 that is:
 \bea
 && \hbox{when} \ \ r_2=r_3 \ \  \
 W_1
 \left| \begin{array}{l}
 \\  \ \  (r_1,c_1,c_1)     \\ g=1
 \end{array}  \right.
 =
 W_1
 \left| \begin{array}{l}
 \\  \ \   (r_1)    \\ g=0
 \end{array}  \right.
  =  W (r_1)   \nonumber \\
 &&  \label{a42} \\
 && \hbox{when} \ \ r_2=r_1 \ \  \
 W_3
 \left| \begin{array}{l}
 \\   \ \   (c_1,c_1,r_3)    \\ g=1
 \end{array}  \right.
 =
 W_3
 \left| \begin{array}{l}
 \\   \ \  (r_3)    \\ g=0
 \end{array}  \right.
 =  W (r_3)   \nonumber
 \eea

 Here following the ideology of [14,15] we have placed the coalesced invariants
 into the breaking point $c_1$ which in a natural way marks the domain of
definition
 of the sought solution with respect to the variables $r_1$ and $r_3$, namely
(see Fig. 2b)
 \be
 r_1 \le c_1 \le r_3   \label{a44}
 \ee
 Using arbitrarity of the parameter $\zeta$ in ~(\ref{a41}) we present the
 boundary conditions for the desired $W_j$'s in the final form
 \bea
 && W_1 (r_1, c_1, c_1, c_1, c_1) = W (r_1) \ , \nonumber \\
 && \label{a45} \\
 && W_3 (c_1, c_1, r_3, r_3, r_3) = W (r_3) \ , \nonumber
 \eea
 An analogous consideration for the left-hand single-phase wave gives
 two more conditions
 \bea
 && W_3 (r_3, r_3, r_3, c_2, c_2) = W (r_3) \ , \nonumber \\
 && \label{a46} \\
 && W_5 (c_2, c_2, c_2, c_2, r_5) = W (r_5) \ , \nonumber
 \eea
 and, obviously,
 \be
 r_3 \le c_2 \le r_5 \ .   \label{a47}
 \ee
 So we have four $(2g)$ boundary conditions of the Goursat type
 for the system ~(\ref{a12}) which is, as it was mentioned above,
 specified by five $(2g+1)$ functions of one variable. The rest
 condition is the boundedness of the $W_j$'s in all the domain
 of definition which provides the boundedness of the hodograph
 solution ~(\ref{a12}) at finite $x,t$. The limited solution is
 given by formulas ~(\ref{a25a}), ~(\ref{a34}) and is parametrized
 precisely by $2g$ arbitrary functions of one variable. Substituting
 this solution  into the boundary conditions
 ~(\ref{a45}), ~(\ref{a46}) one arrives after some calculations at four linear
Abel
 equations resolving of which gives a simple representation for the
 unknown functions $\psi _j (\l)$ in (35).
 \bea
 && \psi _{2k-1} (\l) = \frac{(-1)^k}{2\pi} \intl _{\l} ^{c_k}
 \frac{W(x)}{\sqrt{x-\l}} dx \ ,  \nonumber \\
 && \label{a48} \\
 && \psi _{2k} (\l) = \frac{(-1)^k}{2\pi} \intl ^{\l} _{c_k}
 \frac{W(x)}{\sqrt{\l-x}} dx \ ,  \ \ \ k = 1,2. \nonumber
 \eea
 It should be noted that the arbitrary parameters $\xi_k$ in
 ~(\ref{a34}) have been chosen as follows (compare ~(\ref{a35}) with
 ~(\ref{a44}), ~(\ref{a47}) ):
 \be
 \xi _1 = c_1 \ , \ \ \xi_2 = c_2  \label{a49}
 \ee
 to satisfy the conditions ~(\ref{a35}).

 Examination of the general case of evolution of the initial perturbation
 with $g$ breaking points is completely analogous and does not
 involve any additional difficulties. The result is: the representation
 ~(\ref{a48}) is valid for any $g$. Thus, the arbitrary functions $\psi _k(\l)$
 in the general limited solution ~(\ref {a34}) are the Abel transformations
 of the parts of the inverse to the initial perturbation function $W(r)$
 which are divided by the breaking points. We emphasize that this result is
 valid only for monotonic initial data which have the one-valued
 inverse function. Evolution of the hump-like or large-scale oscillating
 initial perturbation requires an additional consideration
 which will be accomplished in the following papers. Appropriate results
 for the single-phase case can be found in Refs.[30,31].

 So we have found the solution describing the motion of the multivalued
 curve $r_j(x,t)$ proceeding from the single-phase matching. Let us
 examine the initial value problem for the Whitham system (8) corresponding to
the solved
 boundary problem. Making the consequent phase transitions like ~(\ref{a41})
 in the hodograph solution  ~(\ref{a11}) we arrive putting the arbitrary
 parameter $\zeta$ equal to $r_j$ at the system
 \be
 x-V_j (r_j,...,r_j) t = W_j (r_j,...,r_j)   \label{a50}
 \ee
 which is the system of the implicit solutions of  $2g+1$ Hopf
 equations (zero-phase averaged KdV ) with the same initial data
 \be
 x - 6r_j t = W (r_j) \ \ \ j=1,...,2g+1.  \label{a51}
 \ee
 Therefore, both initial data and their smooth evolution are given on the
 pure diagonal $r_1=r_2=...=r_{2g+1}$ in the real $r$-space. By this
 means the desired initial value problem for the Whitham-KdV system has
 the form
 \be
 r_1 (x,0) = r_2 (x,0) =...= r_{2g+1} (x,0) = u_0 (x)  \ . \label{a52}
 \ee
 It should be noted that our consideration really did not require any
 assumptions of the distinct breakings so the latter can be omitted.

 Now we determine the motion of the phase transition boundaries which are
 the multiple caustics for the obtained solution. Really, it follows from
 the potential representation ~(\ref{a9}) that $V_{2k} \br$ coalesces
 with either $V_{2k-1} \br $ or $V_{2k+1} \br $ when $r_{2k}$ coalesces
 with either $r_{2k-1}$ or $r_{2k+1}$ respectively. By this means the lines
 $x_j(t)$ of the phase transition in the $x,t$-plane have the
 multiple characteristic direction in every point and are given by the
 ordinary equations
 \be
 \frac{d x_j}{d t} = V_{j \; multiple} \br  \label{a53}
 \ee
 which should be  considered together with the hodograph solution
 ~(\ref{a11}), ~(\ref{a25a}), ~(\ref{a34}), ~(\ref{a48}).

 Let us discuss briefly the evolution of the Riemann surface topology
 on the obtained family of solutions. The Riemann surface of genus $g$
 is topologically equivalent to a sphere with $g$ handles. These
 handles evolve both in space and in time and some of them degenerate into
 poles on the phase transition boundaries in $(x,t)$-plane. Tending time
 to zero we arrive at the initial Riemann surface which represents a
 sphere with $g$ singularities (poles). By this means we conclude
 that the topology of the evolving Riemann surface from the very
 beginning is given not only by the KdV equation itself but by the whole
 Cauchy problem formulation.

 \section{The fundamental solution to the initial value problem.}

 We give another representation for the solution to the initial value problem
 for the Whitham-KdV system. Substituting the Abel transformations ~(\ref{a48})
 into the solution ~(\ref{a34}) and interchanging the order of integration
 we find
 \be
 \ot \m = \intl _{r_1} ^{r_{2g+1}} W(x) K \m ({\bf r},x) dx  \label{a54}
 \ee
 where $K \m ({\bf r}, x)$ is the fundamental solution (Riemann kernel function
 [32] ) to the initial value problem ~(\ref{a8}), ~(\ref{a52}):
 \bea
  \nonumber \\
 &&  (-1)^k \intl _{r_{2k-1}} ^{min(r_{2k},x)}
 \frac{d\Omega \m}{\sqrt{x-\l}} \ \ \  \hbox{for}
 \ \ r_{2k-1} \le x \le c_k \ ,  \nonumber \\
2 \pi K \m ({\bf r}, x) = &&     \label{a55} \\
 && (-1)^k \intl _{max(r_{2k},x)} ^{r_{2k+1}}
 \frac{d\Omega \m}{\sqrt{x-\l}} \ \ \  \hbox{for}
 \ \ c_k \le x \le r_{2k+1} \ . \nonumber
 \eea

 \section{Summary and Conclusions}
 The generating function of the uniform commuting flows for the multiphase
 averaged KdV equation (Whitham-KdV hierarchy) has been obtained in a
 direct way as depending on an additional parameter $\l$ solution of the
 linear generalized hodograph equations. It has a potential form and the
 potential (generalized frequency) turns out to be proportional to the
 coefficient  of the holomorphic basis differential on the Riemann
 surface of genus $g$, where $g$ coincides with the number of phases.
 The general solution in hodograph space is constructed as a
 superposition of $2g+1$ contour integrals of this generating function
 multiplied by arbitrary functions of one variable. The physical solutions
 corresponding to the Cauchy problem with
 monotonically decreasing initial data with $g$ breaking points for the
 initial KdV equation are distinguished by the requirement of boundedness
 and depend on $2g$ arbitrary functions. Resolving of the appropriate boundary
 problem to the Whitham equations in hodograph space shows that these functions
 are the
 linear Abel transformations of the parts of the initial curve divided by
 breaking points. Evolution of the Riemann surface topology on the obtained
 family of solutions is discussed and it is shown that "initial" Riemann
 surface represents a sphere with $g$ singularities (poles). The
 initial value problem for the Whitham-KdV equations has been formulated
 and its fundamental solution (Riemann kernel function) has been constructed.

 \vspace{0.7cm}

 {\bf Acknowledgments. }

 I would like to thank A.V. Gurevich, A.L. Krylov, V.V. Khodorovskii and
 K.P. Zybin for many fruitful discussions. I express also acknowledgment to
 S.P. Novikov for a number of important comments given at the Workshop
 "Singular limits of dispersive waves - 2" in Zvenigorod where  part of the
 results was discussed.

 I am grateful to M. Pavlov for many useful remarks and discussions which
 were very significant for accomplishing this work.

 Special thanks to E. Kozerenko for the invention of Fig.3 and her time
 and interest in this investigation.

 The work was made possible in part by grant $\#$ RIE300 from the International
 Science Foundation and Russian Government. It is supported also by a
 Fellowship of INTAS Grant 93-2492 and was carried out within the
 research program of the International Center for Fundamental Physics in
 Moscow.

\newpage

\section*{Figure Captions}

\begin{itemize}
\item{1. Splitting of the $(x,t)$--plane in the zero dispersion
limit.}
\item{2. Riemann invariants behaviour for the case $g=2$ at \ \
a) $t=0$ \ \
b) $t_{br} < t < t_{overlap}$}
\item{3. Domain of definition of the single--phase mode:
$r_{2k-1} \le r_{2k} \le r_{2k+1}$ }
\end{itemize}


\begin{thebibliography}{30}
\bibitem{l1}
P.D.Lax and C.D.Levermore, Comm. Pure Appl. Math. {\bf 36}  253,571,809 (1983).

\bibitem{l2}
S.Venakides, Comm. Pure Appl. Math. {\bf 38}, 125 (1985).

\bibitem{l3}
B.A.Dubrovin, V.B.Matveev and S.P.Novikov, Russian Math. Surveys
 {\bf 31}, 59 (1976).

\bibitem{l4}
B.A.Dubrovin,   Russian Math. Surveys  {\bf 36}, 215 (1981).

\bibitem{l5}
P.D.Lax, C.D.Levermore and S.Venakides, in:
Important Developments in Soliton Theory, A.S.Fokas and V.E.Zakharov,
eds., Springer Series in Nonlinear Dynamics X, 205 (1993)

\bibitem{l6}
D.W.McLaughlin, J.A.Strain, Comm.Pure.Appl.Math. {\bf 47}, 1319 (1994)

\bibitem{l7}
G.B.Whitham, Proc. Roy. Soc. {\bf A283}, 238 (1965).

\bibitem{l8}
 H.Flashka, M.G.Forest and D.W.McLaughlin, Comm. Pure Appl. Math.
{\bf 33}, 739 (1980).

\bibitem{l9}
C.D.Levermore, Comm. Partial Differential Equations {\bf 13}, 495 (1988).

\bibitem{l10}
F.R.Tian, Duke Math. Journ. {\bf 74}, 203 (1994).

\bibitem{l11}
S.P.Tsarev, Soviet Math. Dokl. {\bf 31}, 488 (1985).

\bibitem{l12}
I.M. Krichever, Func. Anal. Appl. {\bf 22}, 200 (1988).

\bibitem{l13}
A.V.Gurevich and L.P.Pitaevskii, Sov.Phys.JETP {\bf38}, 291 (1974).

\bibitem{l14}
A.V.Gurevich, A.L.Krylov and G.A.El, JETP Lett. {\bf 54}, 102 (1991).

\bibitem{l15}
A.V.Gurevich, A.L.Krylov, and G.A.El, Sov. Phys. JETP {\bf 74}, 957 (1992).

\bibitem{l16}
F.R.Tian, Comm. Pure Appl. Math. {\bf 46}, 1093 (1993).

\bibitem{l17}
Solar Wind Seven, eds.E.Marsch and R.Schwenn, Pergamon (1992).

\bibitem{l18}
M.Hoppe, C.T.Russel, L.A.Frank, T.E.Eastman, and E.W.Greenberg,
J.Jeoph.Res. {\bf 86}, 4471 (1981).

\bibitem{l19}
V.R.Kudashev, JETP Lett. {\bf 54}, 175 (1991).

\bibitem{l20}
B.A.Dubrovin and S.P.Novikov, Russian  Math. Surveys {\bf 44}, 35 (1989).

\bibitem{l21}
B.A.Dubrovin, Comm. Math. Phys. {\bf 145}, 195 (1992).

\bibitem{l22}
M.V.Pavlov, Doklady Akad. Nauk {\bf 338}, 317 (1994) [in Russian]
(English translation in Russian Math. Dokl.)

\bibitem{l23}
P.G.Grinevich, in NATO ASI series, Vol.235, Singular limits of
dispersive waves, eds. N.Ercolani, I.Gabitov, D.Levermore and
D.Serre, Plenum, New York, 67 (1994).

\bibitem{l24}
L.P.Eisenhart, Annals of Math. {\bf 120}, 262 (1918).

\bibitem{l25}
V.R.Kudashev and S.E.Sharapov, Teor. Mat. Phys. {\bf 87}, 175 (1991)

\bibitem{l26}
G.V.Potemin, Russian Math. Surveys, {\bf 43}, 252 (1988).

\bibitem{27}
O.Wright, Comm. Pure Appl. Math {\bf 46}, 421 (1993).

\bibitem{l28}
N.G.Mazur, Theor. Math. Phys. (in press).

\bibitem{l29}
V.V.Avilov and S.P.Novikov, Sov. Phys. Dokl. {\bf 32}, 366 (1987).

\bibitem{l30}
A.V.Gurevich, A.L.Krylov, N.G.Mazur and G.A.El, Sov. Phys. Dokl. {\bf 37},
198 (1992).

\bibitem{l31}
A.L.Krylov, V.V.Khodorovskii, G.A.El, JETP Lett. {\bf 56}, 323 (1992).

\bibitem{l32}
 R.Courant, Partial Differential Equations II (New York, 1962).

\end{thebibliography}
\end{document}